# Power Law in a Bounded Range: Estimating the Lower and Upper Bounds from Sample Data

Huan-Xiang Zhou

Department of Chemistry and Department of Physics, University of Illinois at Chicago,

Chicago, IL 60607, USA

E-mail: hzhou43@uic.edu

ABSTRACT

Power law distributions are widely observed in chemical physics, geophysics, biology, and beyond. The independent variable $x$ of these distributions has an obligatory lower bound and in many cases also an upper bound. Estimating these bounds from sample data is notoriously difficult, with a recent method involving $O(N^3)$ operations, where $N$ denotes sample size. Here I develop an approach for estimating the lower and upper bounds that involves $O(N)$ operations. The approach centers on calculating the mean values, $\hat{x}_{\min}$ and $\hat{x}_{\max}$, of the smallest $x$ and the largest $x$ in $N$-point samples. A fit of $\hat{x}_{\min}$ or $\hat{x}_{\max}$ as a function of $N$ yields the estimate for the lower or upper bound. Application to synthetic data demonstrates the accuracy and reliability of this approach.

Many types of events in the natural and social worlds occur with frequencies following a power law distribution:

$$p(x) = Ax^{-\alpha}, \quad x > a, \tag{1}$$

where the exponent $\alpha$ is typically between 1 and 3, $A$ is the normalization constant, and $a$ is the lower bound. Examples include the on and off durations of blinking quantum dots,[1,2] expression levels of different mRNAs,[3,4] daily COVID-19 cases in different countries,[5] frequencies of earthquakes with various magnitudes observed in a given region,[6] and the citations of scientific papers.[7,8] In many cases, the range of the independent variable $x$ has an upper bound, $b$, in addition to the lower bound:

$$p(x) = Ax^{-\alpha}, \quad a < x < b. \tag{2}$$

Examples include the on duration of quantum dots,[2] the number of daily COVID-19 cases,[5] and the magnitude of earthquakes in a given region.[9] Many methods have been developed to estimate the parameters ($\alpha$, $a$, and $b$) of power law distributions from sample data. While fast, robust methods are available for estimating the exponent $\alpha$, estimating the bounds, especially the upper bound $b$, is notoriously challenging, and is the subject of the present paper.

The simplest way to estimate $\alpha$ is as the slope of the relation between $p(x)$ and $x$ on a log-log plot. More robust is a maximum-likelihood estimate.[10] Given a sample of $N$ points $\{x_i\}_{i=1}^N$ that putatively obeys a power law distribution without an upper bound, this estimate is given by

$$\alpha = 1 + N / \sum_{i=1}^{N} \ln(x_i/a). \tag{3}$$

Note that this estimate requires the value of the lower bound $a$. The estimation of $a$ is much more elaborate. In the method proposed by Clauset et al.,[11] the estimate of $a$, denoted as $\hat{a}$, was selected among the sample points $\{x_i\}_{i=1}^N$ to minimize the Kolmogorov-Smirnov distance between the theoretical and empirical cumulative distribution functions (CDFs):

$$KS = \max_{x \geq \hat{a} \,\&\, \in \{x_i\}_{i=1}^N} |\hat{C}(x) - C(x)|. \tag{4}$$

The empirical CDF $\hat{C}(x)$ is determined from the subset of sample points that satisfy $x_i \geq \hat{a}$. The theoretical CDF is

$$C(x) = \int_a^x p(x')dx', \qquad (5)$$

with $a$ and $\alpha$ set to the present estimates. This method is complex because the estimation of $\alpha$ and the estimation of $a$ are coupled. It involves $O(N^2)$ operations and is thus also expensive.

The situation becomes even worse for power law distributions with an upper bound. Often the problem of estimating the bounds $a$ and $b$ is avoided by equating $a$ and $b$ to the smallest and largest data values, respectively, in the sample. This practice can result in significant biases in the estimate for $\alpha$, and corrections to the bounds have been proposed, which extrapolate $\hat{a}$ to below the smallest data value and $\hat{b}$ to above the largest data value.[9, 12, 13] Recently, Olmez et al.[14] extended the method of Clauset et al. to estimate $\alpha$, $a$, and $b$ at the same time. In essence, all pairs of points, say $x_i$ and $x_j$ with $x_i < x_j$, from the sample data were tested as possible estimates for $a$ and $b$ (i.e., $\hat{a} = x_i$ and $\hat{b} = x_j$) and the pair that gives the smallest $KS$ is selected as the final choice for $\hat{a}$ and $\hat{b}$. This method involves $O(N^3)$ operations. Still, the resulting $\hat{a}$ is not lower than the smallest data value and $\hat{b}$ is not higher than the largest data value.

Here I develop a new approach for estimating the lower and upper bounds, each involving $O(N)$ operations. The approach centers on calculating the mean values, $\hat{x}_{\min}$ and $\hat{x}_{\max}$, of the smallest $x$ and the largest $x$ in $N$-point samples. First I give the expressions for the exact $x_{\min}$ and $x_{\max}$.[9, 10, 13] To start, note that the CDF $C(x)$ also represents the probability that a data point is less than $x$. Likewise the complementary CDF,

$$\tilde{C}(x) = 1 - C(x), \qquad (6)$$

is also the probability that a data point is greater than $x$. In the sample of $N$ points, the probability that a particular point is between $x$ and $x + dx$ while all the other $N - 1$ points are greater than $x$ is $p(x)dx \cdot [\tilde{C}(x)]^{N-1}$. Accounting for the fact that any one of the $N$ points could be the smallest value, the probability density for the smallest value to be $x$ is thus

$$\pi(x) = Np(x)[\tilde{C}(x)]^{N-1}. \tag{7}$$

The expected value or mean of the smallest $x$ in $N$-point samples is

$$x_{\min} = \int_a^b x\pi(x)\,dx. \tag{8}$$

Inserting the expression for $p(x)$ with the appropriate normalization factor, Eq (8) becomes

$$x_{\min} = \frac{N(\alpha-1)}{[1-(b/a)^{-\alpha+1}]^N} \int_a^b (x/a)^{-\alpha+1}$$
$$[(x/a)^{-\alpha+1} - (b/a)^{-\alpha+1}]^{N-1} dx. \tag{9}$$

$x_{\min}$ should approach $a$ as the sample size $N$ increases. This trend is the basis of the present approach for estimating $a$. When the upper bound is infinite, the integration can be evaluated to yield

$$x_{\min} = \frac{N(\alpha-1)a}{N(\alpha-1)-1} \tag{10a}$$
$$= \frac{a}{1-B/N}, \tag{10b}$$

where $B = (\alpha-1)^{-1}$. For a finite upper bound, Eq (9) can be transformed to a form more appropriate for numerical integration:

$$x_{\min} = \frac{Nb(\alpha-1)}{[(a/b)^{-\alpha+1}-]^N} \int_{a/b}^1 y^{-\alpha+1}(y^{-\alpha+1}-1)^{N-1}dy, \tag{11}$$

where a substitution $y = x/b$ has been made. The numerical integration was implemented according to the mid-point rule with a step size of $10^{-9}$.

The algorithm for estimating $a$ consists of: (1) finding the smallest value in an $N$-point sample; (2) evaluating the mean smallest value ($\hat{x}_{\min}$) among $M$ independent samples; (3) repeating the above steps for several different sample sizes ($N$) while keeping $M$ fixed; and (4) fitting the dependence of $\hat{x}_{\min}$ on $N$ to a function

$$\hat{x}_{\min} = \frac{\hat{a}}{1-(B/N)^\gamma}, \tag{12}$$

where I have introduced an additional parameter, $\gamma$, to accommodate deviations of the estimated $\hat{x}_{min}$ from the exact $x_{min}$ and also potential effects of the upper bound if present.

Figure 1A displays illustrative results for the power law without an upper bound and with $\alpha = 2$ and $a = 1$. Synthetic samples at $N = 10, 20, 50, 100, 200, 500$, and $1000$ were obtained from a random number generator following the targeted power law. $\hat{x}_{min}$ was obtained as the mean smallest value among $M = 100$ samples at each $N$; the error of $\hat{x}_{min}$ was estimated as $SD/\sqrt{M}$, where SD is the standard deviation of the $M$ smallest values. The deviations of $\hat{x}_{min}$ from the exact $x_{min}$ are mostly within errors. The fit of $\hat{x}_{min}$ to Eq (12) yielded $\hat{a} = 1.0009$, extremely close to the true value of 1. The other two parameters, $B$ and $\gamma$, had values of 0.86 and 0.98, to be compared with the exact value of 1 in both cases. Similar results were obtained at $\alpha = 1.5$ and 2.5, where the fits yielded $\hat{a} = 1.002$ and $1.0006$, respectively. A decrease in the discrepancy between estimated and true $a$ with increasing $\alpha$ is expected because the convergence of $\hat{x}_{min}$ to the limiting value $a$ is faster at higher $\alpha$.

Figure 1B presents the corresponding results for the power law with an upper bound of 100. The exact $x_{min}$ [Eq (11)] is slightly lower than the counterpart for the power law without an upper bound [Eq (10b)]. The estimated $\hat{x}_{min}$ agrees with the exact $x_{min}$ within errors. A fit to Eq (12) yielded $\hat{a} = 1.002, 1.0009$, and $0.9999$, respectively, at $\alpha = 1.5, 2.0$, and $2.5$.

The analogous algorithm for estimating $b$ is presented next. The probability density for the largest value to be $x$ in an $N$-point sample is

$$\Pi(x) = Np(x)[C(x)]^{N-1}. \tag{13}$$

The mean largest $x$ in $N$-point samples is

$$x_{max} = \int_a^b x\Pi(x)\, dx \tag{14a}$$

$$= \frac{Na(\alpha-1)}{[1-(b/a)^{-\alpha+1}]^N} \int_1^{b/a} y^{-\alpha+1}(1-y^{-\alpha+1})^{N-1} dy, \tag{14b}$$

where a substitution $y = x/a$ has been made. Equation (14b) was evaluated numercially according to the mid-point rule with a step size of $10^{-5}$. $x_{max}$ should approach $b$ at increasing $N$. This trend is demonstrated in Fig. 2A-C, which presents results calculated at $b =100$ and 1000 and $\alpha = 1.5, 2.0$, and 2.5. To estimate $b$, the $x_{max}$ values within a two-decade interval of $N$, e.g. $[10^2, 10^4]$, were fit to

$$-\ln x_{max} = -\ln \hat{b} + \frac{D}{1 + (N/E)^\delta}, \qquad (15)$$

where $\hat{b}$, $D$, $E$, and $\delta$ are free parameters.

The resulting estimates for $b$ are collected in Fig. 2D. First let us look at the results when the true value of $b$ is 100. At $\alpha = 1.5$, the estimated $\hat{b}$ is $100.5 \pm 0.2$ using $x_{max}$ results in $N \in [10, 10^3]$, so the discrepancy between $\hat{b}$ and $b$ is already within 1% when the estimate is made within a relatively low range of $N$. (The number after the "±" sign represents the standard error of the fit parameter.) At $\alpha = 2.0$, the estimated $\hat{b}$ is $106 \pm 1$ in the $N$ interval of $[10, 10^3]$; the discrepancy from $b$ reduces to within 1% in the next higher interval $[10^2, 10^4]$, where $\hat{b} = 100.24 \pm 0.08$. At $\alpha = 2.5$, $\hat{b}$ is $135 \pm 5$ in $N \in [10, 10^3]$, reduces to $103.8 \pm 0.9$ in $N \in [10^2, 10^4]$, and finally crosses the 1% discrepancy mark to $100.16 \pm 0.06$ in the $[10^3, 10^5]$ interval.

For the higher $b$ of 1000, $x_{max}$ results in higher intervals of $N$ are required to keep the discrepancy between $\hat{b}$ and $b$ within 1%. At $\alpha = 1.5$, this mark is crossed in $N \in [10^2, 10^4]$. At $\alpha = 2.0$, $\hat{b}$ is $1058 \pm 13$ in the $N$ interval of $[10^2, 10^4]$; the discrepancy from $b$ reduces to within 1% in the next higher interval $[10^3, 10^5]$, where $\hat{b}$ is $1002.4 \pm 0.8$. At $\alpha = 2.5$, $\hat{b}$ is $1973 \pm 149$ in $N \in [10^2, 10^4]$, reduces to $1126 \pm 24$ in $N \in [10^3, 10^5]$, and finally crosses the 1% discrepancy mark to $1009 \pm 2$ in the $N$ interval of $[10^4, 10^6]$.

The algorithm for estimating $b$ from sample data consists of: (1) finding the largest value in an $N$-point sample; (2) evaluating the mean largest value ($\hat{x}_{max}$) among $M$ independent samples; (3) repeating the above steps for sample sizes ($N$) spanning a two-decade interval while keeping $M$ fixed; and (4) fitting the dependence of $\hat{x}_{max}$ on $N$ to Eq (15).

Figure 3A-C presents illustrative results for estimating $b$ from sample data when its true value is 100 and $\alpha$ = 1.5, 2.0, and 2.5. Again, synthetic data were obtained from a random number generator following the targeted power law, and $\hat{x}_{max}$ was estimated as the mean largest value among $M$ = 100 samples. The $\hat{x}_{max}$ values from samples of sizes ranging from 10 to $10^5$ agree with the exact $x_{max}$ to within errors. Consequently the further estimate of $b$ by fitting $\hat{x}_{max}$ to Eq (15) is similar to that using the exact $x_{max}$. One difference is that more $\hat{x}_{max}$ values (e.g., 23 in Fig. 3A-C vs. 7 in Fig. 2A-C) are required in a two-decade interval of $N$ to ensure the reliability of the fit value for $\hat{b}$. The $\hat{b}$ values are collected in Fig. 4D. The discrepancies from the true $b$ are similar to those found using exact $x_{max}$ values (Fig. 3D). At $\alpha$ = 1.5, the estimated $\hat{b}$ is 100.6 ± 0.8, with < 1% error, using $\hat{x}_{max}$ values in $N \in [10, 10^3]$. At $\alpha$ = 2.0, the estimated $\hat{b}$ is 111 ± 6 in the $N$ interval of $[10, 10^3]$; the discrepancy from the true $b$ reduces to within 1% in the next higher interval $[10^2, 10^4]$, where $\hat{b}$ is 99.4 ± 0.4. At $\alpha$ = 2.5, the $N$ interval of $[10, 10^3]$ did not produce a reliable fit value for $\hat{b}$. In a slightly higher interval $[20, 2 \times 10^3]$, the fit yielded $\hat{b}$ = 139 ± 36, still with a high standard error. In the next $N$ interval $[10^2, 10^4]$, $\hat{b}$ is 97 ± 1, with both the discrepancy from the true $b$ and the standard error at a low level. Finally $\hat{b}$ crosses the 1% discrepancy mark to 100.1 ± 0.3 in the $[10^3, 10^5]$ interval.

In conclusion, a simple, $O(N)$ algorithm is presented to estimate the lower and upper bounds of power law distributions. The algorithm entails obtaining the smallest and largest values in sample data and fitting these values as a function of sample size. The estimation of the lower and upper bounds is uncoupled from that of the exponent, and does not require any input for the exponent. The estimated bounds can be outside the ranges of observed data. Lastly the algorithm can be easily extended to other types of distribution functions.

**ACKNOWLEDGEMENTS**

This work was supported by Grant GM118091 from the National Institutes of Health.

**DATA AVAILABILITY**

The data that support the findings of this study are available from the corresponding author upon reasonable request.

**REFERENCES**

[1] M. Kuno, D. P. Fromm, H. F. Hamann, A. Gallagher, and D. J. Nesbitt, J Chem Phys **115**, 1028 (2001).

[2] K. T. Shimizu, R. G. Neuhauser, C. A. Leatherdale, S. A. Empedocles, W. K. Woo, and M. G. Bawendi, Phys Rev B **63**, 205316 (2001).

[3] C. Furusawa and K. Kaneko, Phys Rev Lett **90**, 088102 (2003).

[4] H. R. Ueda, S. Hayashi, S. Matsuyama, T. Yomo, S. Hashimoto, S. A. Kay, J. B. Hogenesch, and M. Iino, Proc Natl Acad Sci U S A **101**, 3765 (2004).

[5] B. Blasius, Chaos **30**, 093123 (2020).

[6] B. Gutenberg and C. F. Richter, Bull Seismol Soc Am **34**, 185 (1944).

[7] D. J. d. S. Price, Science **149**, 510 (1965).

[8] G. J. Peterson, S. Pressé, and K. A. Dill, Pro Natl Acad Sci U S A **107**, 16023 (2010).

[9] A. Kijko and M. Singh, Acta Geophys **59**, 674 (2011).

[10] M. E. J. Newman, Contemp Phys **46**, 323 (2005).

[11] A. Clauset, C. R. Shalizi, and M. E. J. Newman, SIAM Review **51**, 661 (2009).

[12] T. Maschberger and P. Kroupa, Mon Not R Astron Soc **395**, 931 (2009).

[13] J. Zhang, Statistics **47**, 792 (2013).

[14] F. Olmez, P. R. Kramer, J. Fricks, D. R. Schmidt, and J. Best, J Stat Comput Simul **91**, 1524 (2021).

**FIGURE CAPTIONS**

FIG. 1. $\hat{x}_{min}$ compared to exact $x_{min}$ and fit to Eq (12) to yield an estimate for the lower bound $a$. (A) Power law distribution without an upper bound; $\alpha = 2$ and $a = 1$. (B) Corresponding results for a power law distribution with an upper bound $b = 100$.

FIG. 2. Fit of the exact $x_{max}$ to Eq (15) to yield an estimate for the upper bound $b$. (A) $\alpha = 1.5$. Fit1, Fit2, Fit3, and Fit4 used data in the following ranges of $N$: [10, $10^3$], [$10^2$, $10^4$], [$10^2$, $10^4$], and [$10^3$, $10^5$], respectively. (B) $\alpha = 2$. Fit1, Fit2, Fit3, and Fit4 used data in the following ranges of $N$: [10, $10^3$], [$10^2$, $10^4$], [$10^2$, $10^4$], and [$10^3$, $10^5$], respectively. (C) $\alpha = 2.5$. Fit1, Fit2, Fit3, and Fit4 used data in the following ranges of $N$: [10, $10^3$], [$10^3$, $10^5$], [$10^2$, $10^4$], and [$10^4$, $10^6$], respectively. (D) Estimated $b$. The upper limits of two-decade $N$ intervals used for estimating $b$ are shown in the legend.

FIG. 3. Fit of $\hat{x}_{max}$ to Eq (15) to yield an estimate for the upper bound $b$. The true $b = 100$. (A) $\alpha = 1.5$. Fit1 and Fit2 used data in the $N$ ranges of [10, $10^3$] and [$10^2$, $10^4$], respectively. (B) $\alpha = 2$. Fit1 and Fit2 used data in the $N$ ranges of [10, $10^3$] and [$10^2$, $10^4$], respectively. (C) $\alpha = 2.5$. Fit1 and Fit2 used data in the $N$ ranges of [20, $2 \times 10^3$] and [$10^3$, $10^5$], respectively. (D) Estimated $b$. The upper limits of two-decade $N$ intervals used for estimating $b$ are shown in the legend.

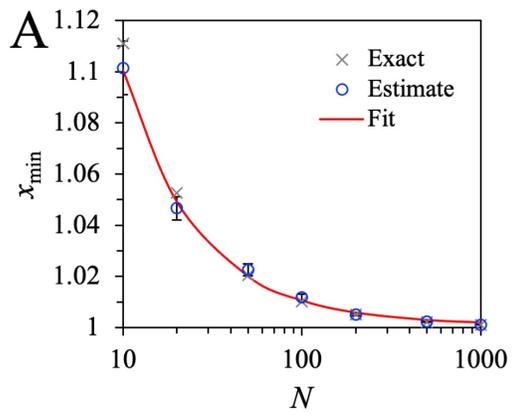 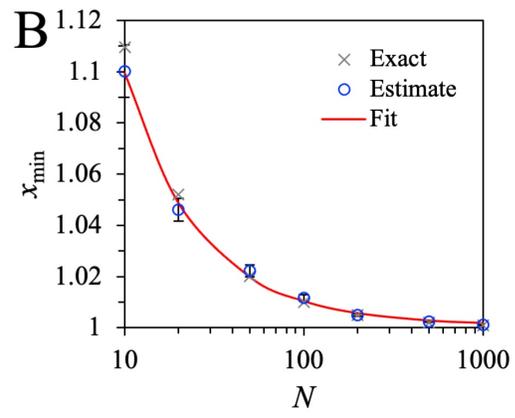

Figure 1

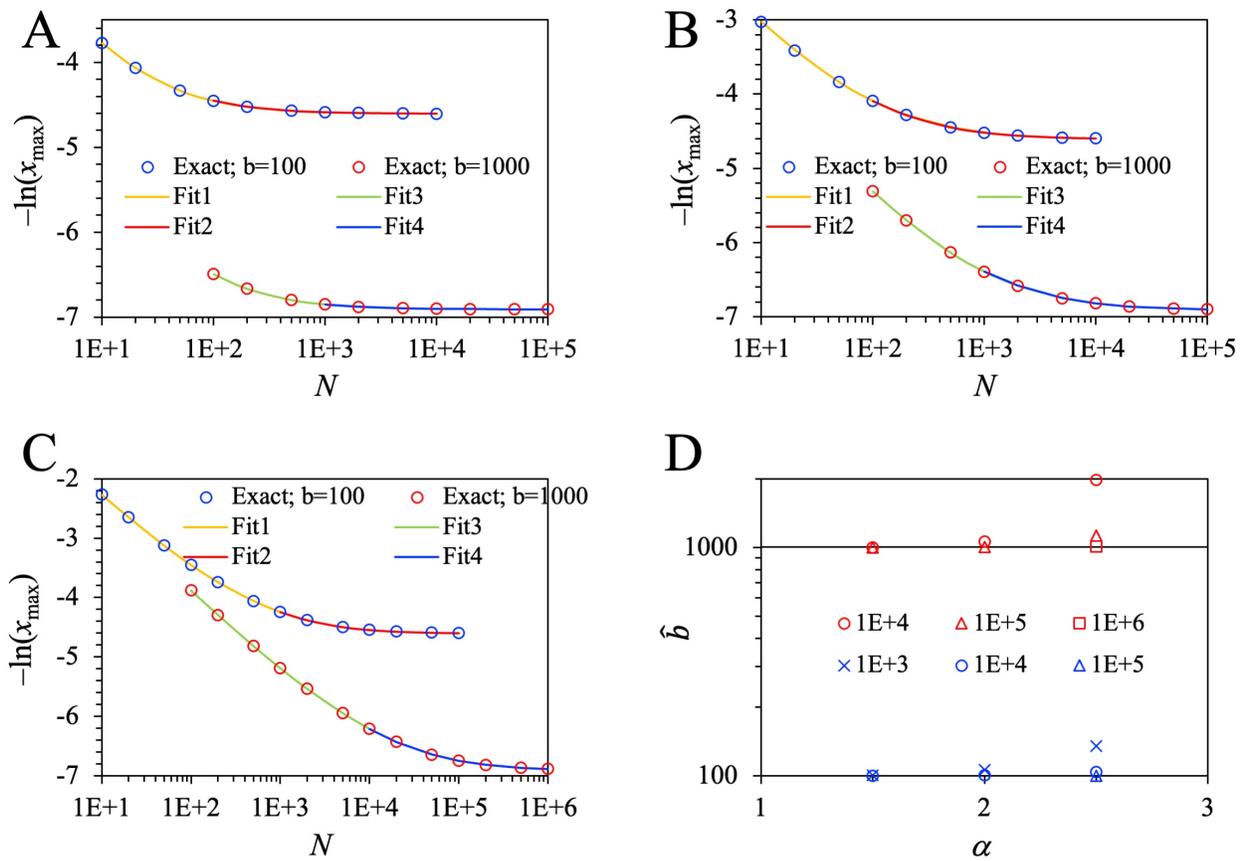

Figure 2

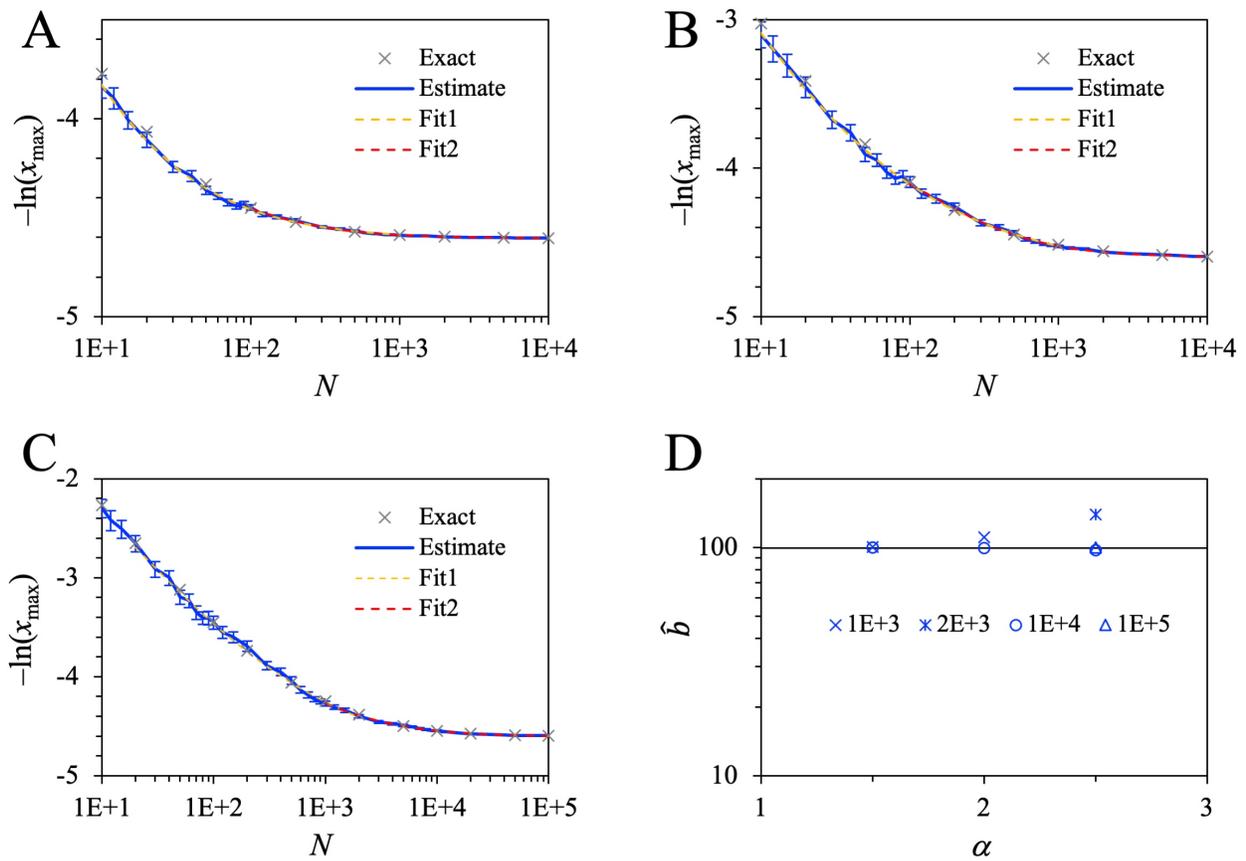

Figure 3